\documentstyle[preprint,aps]{revtex}

\begin{document}

\draft
\preprint{LBL-35981}

\title{How Disoriented Chiral Condensates Form: Quenching
  vs. Annealing
}

\author{Masayuki Asakawa$^a$, Zheng Huang$^b$, and Xin-Nian Wang$^a$}

\address{{$^a$Nuclear Science Division, MS 70A-3307, \\
\baselineskip=12pt Lawrence Berkeley Laboratory, Berkeley, CA 94720}\\
{$^b$Theoretical Physics Group, MS 50A-3115, \\
\baselineskip=12pt Lawrence Berkeley Laboratory, Berkeley, CA 94720}}

\date{August 15, 1994}
\maketitle
\begin{abstract}

We demonstrate that semiclassical fluctuations, their relaxation,
and the chiral phase transition are automatically incorporated
in the numerical simulations of the classical equations of motion
in the linear $\sigma$-model when longitudinal
and transverse expansions are included. We find that
domains of disoriented chiral condensate with 4--5 fm in size
can form through a quench while an annealing leads
to domains of smaller sizes. We also demonstrate that quenching
cannot be achieved by relaxing a chirally symmetric system through
expansion.
\end{abstract}


\narrowtext

One of the proposed explanations for the Centauro events \cite{lfh80}
in high energy cosmic ray experiments is the coherent emission
of pions from a large domain of disoriented chiral condensate
(DCC) \cite{ar89}. However, if the system has to go through
an equilibrium phase transition, quark masses, though much smaller
than the intrinsic QCD scale, prevent DCC domains from
reaching a size larger than $1/T_c$ \cite{rw92}.
As an alternative, Rajagopal and Wilczek \cite{rw93}
proposed that a nonequilibrium phase transition through
quenching can generate large DCC domains. Their
numerical simulation indeed observed the amplification of the
long wavelength modes of the pion fields, but did not shed light
on the size of DCC domains. Similar simulations
by Gavin, Gocksch, and Pisarski \cite{ggp} are, however,
not conclusive about the exact size of DCC domains, since
they only looked at the pion fields averaged over the transverse
dimensions which cannot reveal domains smaller than the lattice
size.

To model a quench, Rajagopal and Wilczek argued that one
can evolve the classical fields according to the zero
temperature equations of motion from a chirally symmetrical
initial condition with short correlation lengths.
In this Letter, we shall argue that semiclassical fluctuations
introduced in the initial configuration actually render the
effective potential to a non-zero temperature one. The interaction
between the mean fields and the semiclassical fluctuations as
well as their evolution can be automatically included in the
numerical simulations of the equations of motion.
Using an ensemble averaging technique, we demonstrate the
relaxation of the fluctuations and the occurrence of the
chiral phase transition due to the longitudinal \cite{hw94} and
transverse expansions which are consistently included in our
study, whereas in earlier simulations \cite{rw93,ggp}
the system only evolves in an approximately constant \cite{note1}
but non-zero temperature effective potential. By choosing different
initial configurations (both the mean fields and
the fluctuations), we study the evolution of the system
in both quenching\cite{rw93} and annealing\cite{gm94,boy}
scenarios.

In the standard linear $\sigma$-model, the equations of
motion are given by,
\begin{equation}
\left ( \frac{\partial ^2}{\partial t^2} - \nabla ^2 \right ) \phi
= \lambda(v^2 - \phi^2 )\phi  +H n_{\sigma},\label{em}
\end{equation}
where $\phi \equiv (\sigma, \mbox{\boldmath $\pi$})$ is a vector
in internal space, $n_{\sigma}=(1,\mbox{\bf 0})$, and
$Hn_{\sigma}$ is an explicit chiral symmetry breaking term
due to finite quark masses.
In the following, we shall use $\lambda=19.97$, $v=87.4$ MeV,
and $H= (119~{\rm MeV})^3$, with which $m_\pi = 135$ MeV
and $m_\sigma = 600$ MeV.
We emphasize that the $\phi$ fields in the above equations
can include both the mean fields and the semiclassical fluctuations.
Thus, we can separate $\phi$ into two parts,
\begin{equation}
\phi = \langle\phi\rangle + \delta\phi, \label{avfl}
\end{equation}
where $\langle\phi\rangle$ are the mean
fields and  $\delta\phi$ are the semiclassical fluctuations
around $\langle\phi\rangle$. Using Eq. (\ref{avfl}) and
taking the average of Eq. (\ref{em}),
we obtain the equations of motion for the mean fields in
the Hartree approximation \cite{gm94},
\begin{equation}
\left ( \frac{\partial ^2}{\partial t ^2} - \nabla ^2 \right )
\langle\phi\rangle
 =  \lambda(v^2 - \langle\phi\rangle^2 -
3\langle\delta\phi_{\|}^2\rangle -\langle\delta\phi_\bot^2 \rangle )
\langle\phi\rangle  +H n_{\sigma}, \label{hffl}
\end{equation}
where $\langle\phi\rangle^2=\langle\phi_i\rangle\langle\phi_i\rangle$,
$\delta\phi_{\|}$ is the component of the fluctuation
parallel to $\langle\phi\rangle$,
and $\delta\phi_{\bot}$ is the orthogonal component.
These equations imply that the motion of the mean
fields is determined by an effective potential,
\begin{equation}\label{effpot}
V(\langle\phi\rangle)=\frac{\lambda}{4}
(\langle\phi\rangle^2  + 3 \langle\delta\phi_{\|}^2\rangle
+\langle\delta\phi_\bot^2 \rangle-v^2 )^2
-H\langle\sigma\rangle,
\end{equation}
which clearly differs from the zero temperature one in the
presence of the fluctuations. By varying the
level of fluctuations, chiral symmetry can be restored
or spontaneously broken. The above effective potential is
very generic since no assumption has been made for the
fluctuations except that they are only of a classical nature.
If the fluctuation terms in Eq. (\ref{effpot})
are replaced by their counterparts in a finite temperature
field theory, the well-known one loop effective potential
at finite temperature \cite{dj74} is recovered. Therefore,
for the mean fields, and as we shall show for the fluctuation
fields as well, the classical equations of
motion have already included the effect
of fluctuations present in the effective potential.
This might look surprising, but can be easily understood in
a thermal equilibrium case. In a finite temperature
field theory, the temperature dependence of the
effective potential arises from the  on-shell
part of the propagator. Since no contribution from
virtual particles is involved, all thermal corrections
at one loop level are purely classical.

Since we shall neglect the corrections due to quantum
fluctuations \cite{boy,kluger}, we can also derive the
equations of motion for the fluctuation
fields $\delta\phi$ by subtracting Eq. (\ref{hffl}) from
Eq. (\ref{em}). These two sets of coupled equations
for $\langle \phi\rangle$ and $\delta\phi$ can,
therefore, be consistently solved through
numerical simulations  \cite{rw93,ggp} of the
classical equations of motion.
Since the fluctuations and their squared ensemble
averages can evolve with time according to a given relaxation
mechanism \cite{hw94,gm94,boy}, the time evolution
of the field configuration obtained
from Eq. (\ref{em}), already includes the effect of the time
dependence of the effective potential.
The use of equations of motion does not ensure that
the effective potential takes its zero temperature form or
that chiral symmetry is spontaneously broken.
What matters most is the initial fluctuation of the system.

When $\delta^2\equiv(3\langle\delta\phi_{\|}^2\rangle+
\langle\delta\phi_\bot^2\rangle)/6$ is large enough,
chiral symmetry is restored (approximately, due to $H\neq 0$).
If the explicit chiral
symmetry breaking term is neglected, the phase transition
takes place at the critical fluctuation, $\delta^2_c\equiv v^2/6$.
For $\delta^2<\delta_c^2$, the effective potential takes
its minimum value at $\langle\phi\rangle=(\sigma_e,\mbox{\bf 0})$,
where $\sigma_e$ depends on $\delta^2$. When the mean fields
are displaced from this equilibrium point to the central lump
of the ``Mexican hat'' ($\langle\phi_i\rangle\sim 0$),
modes below the critical momentum,
\begin{equation}\label{crmom}
k_{c} = \left [ \lambda ( v^2 - \langle\phi\rangle^2 -
3 \langle\delta\phi_{\|}^2\rangle
-\langle\delta\phi_\bot^2 \rangle ) \right ]^{1/2},
\end{equation}
become unstable and thus DCC
domains can form. Since the domain size is directly
related to the time scale during which these
modes are unstable, it strongly depends on the initial
condition, $\delta^2$ and $\langle\phi\rangle$ of the system.

Let us now consider three different scenarios. (i) In a
quenching scenario, the initial fluctuation is below the
critical value, $\delta^2<\delta_c^2$,
and $\langle\phi_i\rangle\sim 0$. As the mean fields
roll down from the central lump of the potential,
pion modes below $k_c$ will be amplified. In the meantime,
as the system cools down and the fluctuation decreases via, e.g.,
an expansion, the effective potential will also
change and the equilibrium point of the potential
$(\sigma_e,\mbox{\bf 0})$ moves towards
the zero temperature value $(f_{\pi},\mbox{\bf 0})$.
This will increase the roll-down time and lead to
a larger domain size. (ii) On the other hand, if we
initially choose $\langle\phi\rangle$ to be very
close to the equilibrium point of the effective
potential, $(\sigma_e,\mbox{\bf 0})$, the mean
fields and the effective potential may both evolve
so that the system always oscillates around the
equilibrium position. We refer to this scenario as
a cold annealing since the system starts with an
effective potential in which chiral symmetry is spontaneously
broken. (iii) What we shall call a hot annealing scenario
is similar to the cold annealing except that the initial
fluctuation is much larger than the critical value,
$\delta^2\gg\delta^2_c$, so that chiral symmetry is
almost restored. In both annealing cases, the mean fields
can evolve almost synchronously with the effective potential
so that the system only oscillates around the equilibrium
point, i.e., $ \langle\phi\rangle^2 \approx v^2
 - 3 \langle\delta\phi_{\|}^2\rangle - \langle\delta\phi_\bot^2 \rangle $.
One can expect then that the low momentum modes
are less amplified and the domain size is smaller
than in the quenching case.

In order to confirm our arguments, we have carried out
numerical simulations of Eq. (\ref{em}) including both the longitudinal
and transverse expansions. We assume boost invariance
in the longitudinal direction so that the longitudinal
expansion is automatically included. To consider the
transverse expansion, we use a cylindrical
boundary condition. The initial $\phi$ fields
are randomly distributed according to a Gaussian form
with the following parameters:
\begin{eqnarray}
\langle\sigma\rangle & = & (1-f(r))(f_\pi-\sigma_0)+\sigma_0 \nonumber \\
\langle\pi_i \rangle & = & 0, \nonumber \\
\langle\sigma^2 \rangle -\langle\sigma\rangle^2
& = & \langle\pi_i^2 \rangle = \delta^2_0 f(r), \nonumber \\
\langle\dot{\sigma}\rangle & = & \langle\dot{\pi_i} \rangle =0, \nonumber \\
\langle\dot{\sigma}^2 \rangle & = & \langle\dot{\pi_i}^2 \rangle
= 4\delta^2_0 f(r),
\label{initial}
\end{eqnarray}
where $r = (x^2 + y^2)^{1/2}$ is the radial coordinate, the
dot stands for the derivative with respect to the proper
time $\tau = (t^2 - z^2 )^{1/2}$, $\sigma_0$ and $\delta^2_0$
are constants which we can vary for different scenarios.
We have introduced an interpolation function,
\begin{equation}
f(r)=\left[\exp\left( \frac{r - R_0}{\Gamma} \right)+1\right]^{-1},
\end{equation}
to describe the boundary condition. $R_0$ is the radius of
the initially excited region where fluctuations exist and
the mean fields are different from their vacuum expectation
values. Outside this region, the vacuum configuration,
$\phi=(f_{\pi},\mbox{\bf 0})$, is imposed.
$\Gamma$ is the thickness of the transient region.
The results presented in this Letter are obtained with
$R_0=5$ fm and $\Gamma= 0.5$ fm.

In our numerical simulations, we have used two-step Lax Wendroff method,
a version of the leap frog method \cite{full}.
We have used an initial correlation length, $\ell_{\rm corr}= 0.5$ fm.
Usually the spacing $a$ of the lattice on which numerical
simulations are carried out has been identified
to $\ell_{\rm corr}$ \cite{rw93,ggp}. To include the initial
correlation and to reduce the finite size effect, we have
adopted a lattice spacing smaller than $\ell_{\rm corr}$ \cite{note2}.
The initial fields are therefore uniform within
$\ell_{\rm corr} \times \ell_{\rm corr}$ squares.
Since domain formation is caused by the amplification of
low momentum modes, the domain size should not be very sensitive
to the value of $\ell_{\rm corr}$. We have actually confirmed this
in our numerical simulations \cite{full}.

For the three scenarios we consider here, we take (i)
$\delta_0^2=\delta_c^2=v^2/6$, with which the
system is about to go through a phase transition. The
initial field configuration is set to $\sigma_0=0$
for the quenching case. (ii) For a cold annealing case, we take
$\sigma_0=\sigma_e=44$ MeV, in order to have an equilibrium
initial configuration with the given fluctuation.
(iii) For a hot annealing case, we take $\delta^2_0=v^2/4$
and $\sigma_0=\sigma_e=20$ MeV.

We define a correlation function $C(r,\tau)$ as
\begin{equation}
C(r,\tau)=
\frac{\sum_{i,j}\mbox{\boldmath $\pi$}(i)\cdot
\mbox{\boldmath $\pi$}(j)}
{\sum_{i,j}|\mbox{\boldmath $\pi$}(i)| |\mbox{\boldmath $\pi$}(j)|},
\end{equation}
where the sum is taken over those grid points
$i$ and $j$ such that the distance between $i$
and $j$ is $r$.
In Fig. 1(a), we show the time evolution of the
correlation function for the quenching case.
Throughout the calculations shown in this Letter,
we take the initial time $\tau_0 = 1$ fm and a
lattice spacing $a=0.25$ fm.
At $\tau = \tau_0$, there is no correlation beyond
the initial correlation length $\ell_{\rm corr}$.
We can clearly see that a long range correlation
emerges at later times.
Note that the apparent shrinking of correlation length
at $\tau =7$ fm is a little misleading.
Since the resultant pion distributions only depend on
$\pi_i^2$, $C(r, \tau)<0$
should be regarded as the manifestation of the
correlation as long as $C(r, \tau)\not\approx 0$.
At $\tau=7$ fm, the typical correlation length
is as long as $r\sim 2.5$ fm. In Fig. 1(b), we compare the
results of the quenching, hot and cold annealing scenarios
at $\tau = 7$ fm.  We observe that quenching gives the
largest correlation among the three cases. We have also checked
that changing $\sigma_0$ to 0 in the hot annealing case
does not help much to create larger domains, since
the system has moved to the equilibrium position before the
chiral phase transition takes place. In other words,
a quenching condition can never be realized through
a hot annealing.  The situation
does not improve much even if a second order phase transition
is assumed ($H=0$), because the expansion time scale is
too short for any long range correlation to develop.
In the quenching case, the expansion of the system
reduces the fluctuation, and as a result, the
evolving effective potential provides a longer
roll-down time for the system to form larger domains.
We have in fact checked that for larger $\tau_0$ cases,
where relative longitudinal expansion is slower,
a smaller correlation is generated \cite{full}.

In Fig. 2, we show the time evolution of the average
$\sigma$ field, $\langle\sigma\rangle$, and the average
fluctuations, $\langle\delta\sigma^2\rangle^{1/2}$ and
$\langle\delta\pi_1^2\rangle^{1/2}$.
$\langle\delta\pi_2^2\rangle^{1/2}$ and
$\langle\delta\pi_3^2\rangle^{1/2}$ are similar to
$\langle\delta\pi_1^2\rangle^{1/2}$. In this calculation,
the initial fluctuation, $\delta^2_0 = 3v^2/8$, has been
taken to be much larger than the critical value so that
chiral symmetry is almost restored. For $\sigma_0$, we have
taken its value to be $\sigma_e=9$ MeV. We have used 100 events and
averaged over the central region $r\leq 3$ fm. We see that
$\langle\delta\sigma^2\rangle$ and $\langle\delta\pi_1^2\rangle$
decrease on the average with time due to the longitudinal and
transverse expansion.  On the other hand, $\langle\sigma\rangle$
increases, following the equilibrium position of the
evolving effective potential. In principle, $\langle\sigma\rangle$
approaches to its vacuum value, $f_{\pi}$, as $\delta^2$
goes to zero. A very interesting and
important point is that $\langle\delta\sigma^2\rangle$
and $\langle\delta\pi_1^2\rangle$ decouple from each other
at about $\delta=35$ MeV, which is about the value of the
critical fluctuation, $\delta_c=(v^2/6)^{1/2}$.
This decoupling is nothing but the manifestation of
the mass splitting during the chiral phase transition:
the pion mass becomes
smaller and the sigma mass becomes larger.
Obviously, the semiclassical fluctuations also
experience an evolving effective potential as well as
the mean fields ( as seen in Fig. 2, $\langle\sigma\rangle$
rapidly increases after the phase transition as the system
follows the ever-changing equilibrium position).

Finally, to demonstrate domain formation, we show the contour plot
of $\pi_2$ in Fig. 3 for one event in our quenching case as
a function of the transverse coordinates at the initial
time $\tau=\tau_0=1$ fm and $\tau=5$ fm. We can clearly see two
large domains with opposite signs in $\pi_2$ at $\tau=5$ fm in this event,
whereas initially there exists no structure inside the chirally
restored region. We also note an apparent transverse expansion and
decreasing fluctuations in the inner region.
The domain formation is more dramatic for a smaller system \cite{full}.

In summary, we have shown that the usual prescription for a quench
actually already includes the effect of fluctuations.
The relaxation of the semiclassical fluctuations and their effect
on the effective potential are automatically included
in the evolution of the system which undergoes both longitudinal
and transverse expansions. Mass splitting of the pion and
sigma fields at the classical level during the chiral phase
transition has been clearly demonstrated.
We have also shown in our numerical calculations
that DCC domains of a typical size up to 4-5 fm can form
for realistic parameters in the linear $\sigma$-model if the quenching
initial condition is realized. Furthermore, we have demonstrated that
such a quenching condition cannot be achieved by relaxing
a system from a chirally symmetrical phase through expansions.
We will discuss elsewhere \cite{full} whether
and how a quenching initial state can be realized in hadronic
or nuclear collisions.

\acknowledgments
We would like to thank S. Gavin and B. M\"uller for discussions.
This work was supported by the Director, Office of Energy Research,
Office of High Energy and Nuclear Physics, Divisions of High Energy Physics
and Nuclear Physics of the U.S. Department of Energy under Contract
No. DE-AC03-76SF00098. Z. H. was also supported by the Natural Sciences and
Engineering Research Council of Canada.

\newpage
\centerline{\bf Figure Captions}
\vskip 15pt
\begin{description}
\item[Fig. 1] Correlation functions (a) at $\tau=1$ fm (initial state),
              3 fm, and 7 fm for the quenching initial condition;
              (b) for the quenching, hot and cold annealing scenario
               at $\tau=7$ fm.
\item[Fig. 2] Evolution of $\langle\sigma\rangle$,
              $\langle\delta\sigma^2\rangle^{1/2}$, and
              $\langle\delta\pi_1^2\rangle^{1/2}$ as functions
              of $\tau$. The initial fluctuation is taken to be,
              $\delta^2 = 3v^2/8$, and $\sigma_0 = 9$ MeV.
              The averages are made over 100 events and within $r \leq 3$ fm.
\item[Fig. 3] Contour plot of $\pi_2$ field in an event
              at $\tau=\tau_0=1$ fm and $\tau = 5$ fm. A quenching initial
              condition is used as in Fig. 1. The $\pi_2$ field
              has opposite signs in the two large domains.
\end{description}


\begin{references}
\bibitem{note1} Due to energy conservation, the evolution of the
        mean fields will affect the fluctuations which
        in turn can change the effective potential slightly.
\end{references}
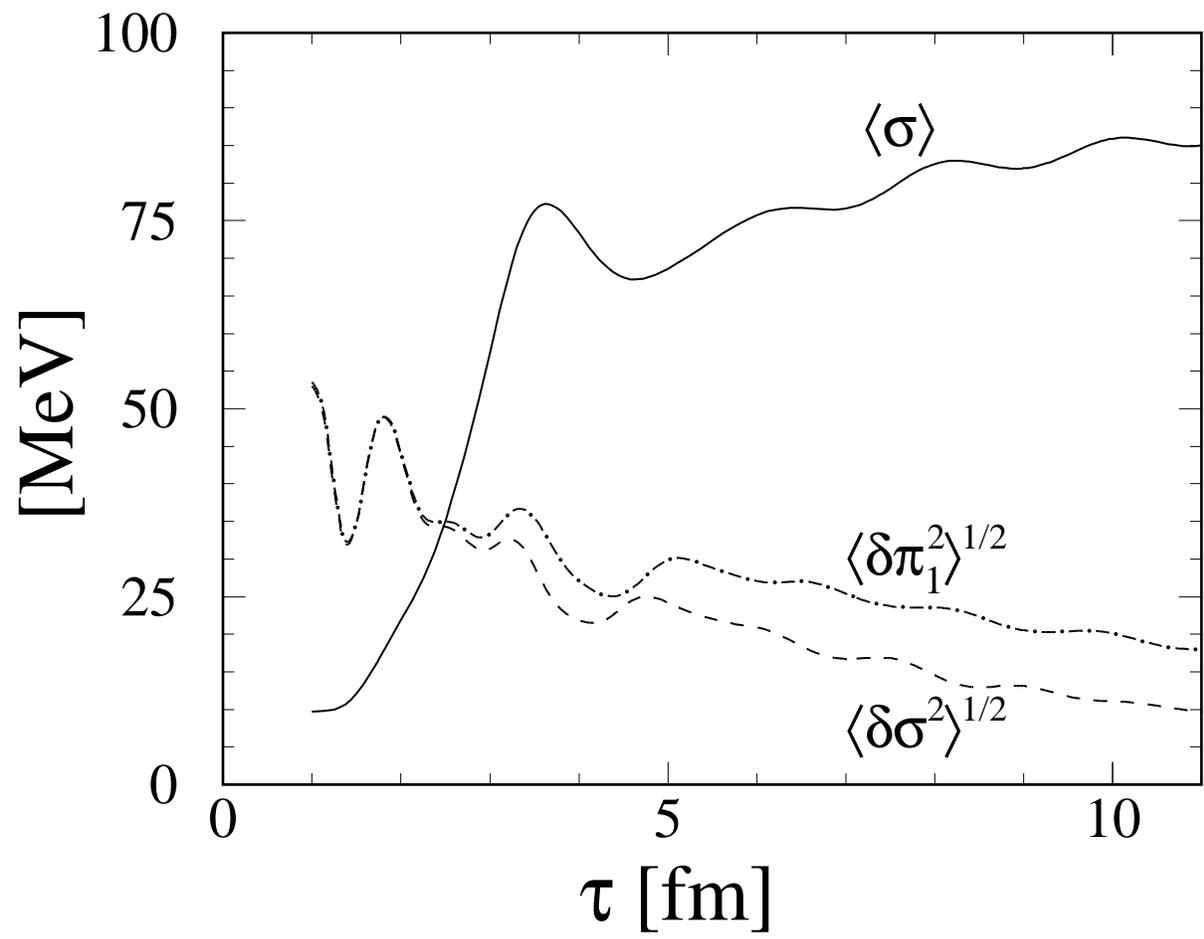
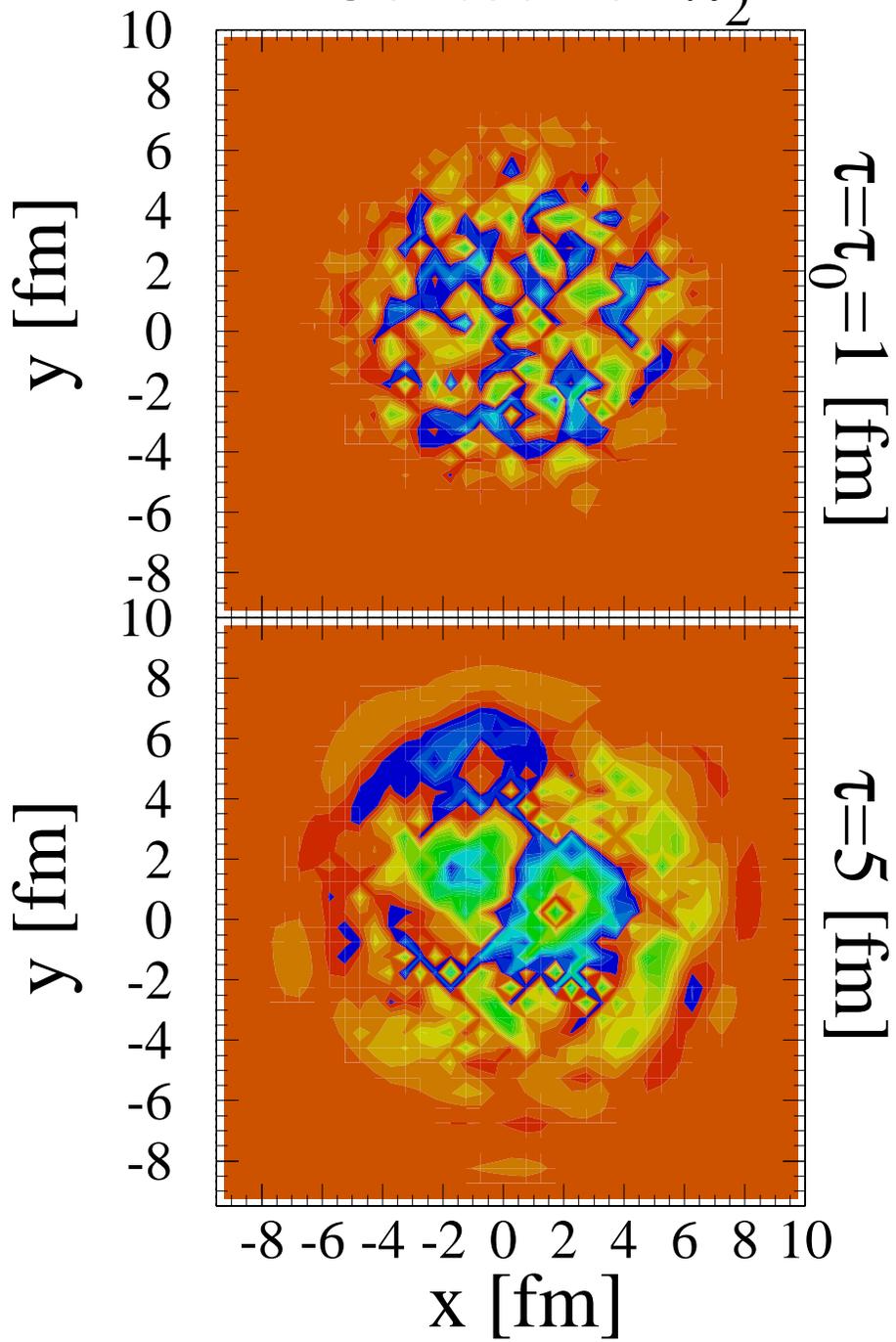
\end{document}